\documentclass[11pt]{article}
\usepackage{amssymb}
\usepackage{amsmath}
\usepackage{amsthm}

\usepackage{graphicx,color}

\newcommand{\ket}[1]{\left | #1 \right \rangle}

\def\openone{\leavevmode\hbox{\small1\kern-3.8pt\normalsize1}}

\theoremstyle{definition}

\begin{document}
\title{\LARGE\bf
Efficient classical verification\\ of quantum computations}

\author{Richard Jozsa and Sergii Strelchuk\\[3mm]
  \small\it DAMTP, Centre for Mathematical Sciences, University of Cambridge,\\ \small\it Wilberforce Road, Cambridge CB3 0WA, U.K.\\[1mm]
  }

\date{}

\maketitle

\begin{abstract}
We propose an efficient scheme for verifying quantum computations in the `high complexity' regime i.e. beyond the remit of classical computers. Previously proposed schemes remarkably provide confidence against arbitrarily malicious adversarial behaviour in the misfunctioning of the quantum computing device. Our scheme is not secure against arbitrarily adversarial behaviour, but may nevertheless be sufficiently acceptable in many practical situations. 
With this concession we gain in manifest simplicity and transparency, and in contrast to previous schemes, our verifier is entirely classical. It is based on the fact that adaptive Clifford circuits on general product state inputs provide universal quantum computation, while the same processes without adaptation are always classically efficiently simulatable.

\end{abstract}

\section{Introduction}\label{intro}
Establishing confidence in the output of a quantum computing device operating in the `high complexity' regime i.e. beyond the remit of classical computers, will be an important issue as our first quantum computers become available. Some previously proposed schemes for such verification \cite{aboe08,av12, bro15, abem17} have been based on adaptations of the formalism of interactive proof systems (IP) from classical complexity theory \cite{aubar}, while others \cite{mckag16, morfit16, hayhad16} have been based on the formalism of measurement based computing (MQC). Verification has also played an important role in the cognate subjects of blind quantum computing \cite{ch05, bfk09}, and self testing and device independent protocols \cite{mmmo06, mckag16, reich13}.

Most of the previously proposed schemes have been designed to provide confidence in correctness of the output against the most general prospective malicious or adversarial behaviour in the misfunctioning of the computer. But although a stance of such extreme guardedness may be appropriate and relevant in fields such as cryptography, information security and financial transaction, it is not normally adopted in standard scientific method for scrutiny of experimental outcomes, as exemplified in the spirit of Einstein's quote: {\em ``Nature hides her secret because of her essential loftiness, but not by means of ruse"}. Indeed in practice essentially all of our physical devices and experimental demonstrations are subject to prospective, but accepted, loopholes, or rest on accepted prior assumptions whose failure could invalidate the claimed demonstrations.

Essentially the same issue of verification arises also in any technological development in computing where a new ``super-computer'' is built, whose utility rests on its ability to outperform previously existing computers. In testing it and developing confidence in the correctness of its operation, we do not normally check it against prospective arbitrarily malicious behaviour but are generally satisfied by extrapolating confidence from a variety of more limited confirmable checks.

The all-encompassing secure level of confidence required in most previous schemes comes at a considerable cost. In the IP schemes verifying a quantum circuit $C$ that solves a desired BQP decision problem, the algorithm $C$ itself is not directly implemented. Instead the structure of $C$ is embedded in a very much more complicated interactive protocol. The extra complication here is needed to give confidence in the output of an untrusted computer and to be able to reliably detect arbitrarily malicious misfunctioning. Although the correctness of the output of any such run is guaranteed (with high probability), there is nevertheless no implied guarantee that the computer could correctly implement the original circuit $C$ by itself, yet this is probably how we would most likely wish to continue to use the computer. Indeed if we were to fully embrace the spirit of the IP formalism as a necessary requirement, we would be taking seriously even the scientifically implausible possibility that a malicious computer would behave correctly for IP protocol runs but then operate with deliberate deception on standard (direct) subsequent runs. As such, the remarkable achievement of unconditional verifiability may yet perhaps be somewhat self-defeating for standard practical use of the computing device. 

The MQC schemes on the other hand are generally simpler and directly applicable to the computation being verified, which must however be implemented in MQC form. Furthermore the verifier generally needs to have the quantum capability of performing 1-qubit measurements on individually held qubits that are still entangled with others or the rest of the computer, which amounts to the verifier having the ability to perform individual computational steps that would suffice for universal quantum computing in this context (which is also the case for the IP scheme of \cite{aboe08, abem17} in the circuit model context). As such, these schemes, while being significant developments in the theory of verification, are perhaps less relevant for issues of realistically efficient verification capability {\em per se}. In fact all previous schemes (except some  \cite{mckag16, reich13} having multiple provers, so not directly relevant for our setting) require the verifier to have some quantum processing capability and a quantum communication channel to the prover.

In contrast to the above, in our scheme the verifier is entirely classical, using only polynomially-bounded classical computing resources and only classical communication with the prover. Furthermore it is based on the prover (i.e. our quantum computing device) simply running the given computational process itself directly. However there are associated limitations too: the confidence in correctness of the output will not be secure against arbitrarily malicious malfunctioning of the computer, but nevertheless we would expect that it could be acceptable in many realistic situations, to an extent that's not dissimilar from commonly accepted scientific practice. Indeed with this concession we gain in manifest simplicity and transparency. In particular (in view of Einstein's quote), by not requiring computational blindness \cite{ch05,bfk09} in our scheme, we gain much simplicity.

Our scheme will be based on the fact that adaptive Clifford computations (on product state inputs that can include 1-qubit magic states) provide universal quantum computation \cite{brkit}, whereas non-adaptive Clifford computations with the same inputs are always classically efficiently simulatable \cite{jvdn}. Intuitively, this will enable us to reduce the the running of a universal quantum computer to a classically simulatable process, {\em after} the machine has completed its run, and hence, with further testing runs of the same sequence of quantum operations that occurred in the initial run, we can efficiently verify that the machine was able to correctly run that sequence.

Our scheme, like some previously proposed ones, is similarly unable to verify outputs for processes involving multi-bit outputs with final measurements on more than $O(\log n)$ lines (where $n$ is the number of lines in the original circuit) i.e. we are unable to verify suitably complex sampling (in contrast to decision) problems. In our case this arises because although adaptive Clifford circuits with arbitrary product state inputs are efficiently classically simulatable for single bit outputs, they are not known to generally remain so for multi-bit outputs on more than $O(\log n)$ lines \cite{jvdn}.

\section{The verification scheme}\label{scheme}

Suppose we have available a quantum computing device that can allegedly perform Clifford gates and computational basis measurements, and we can also reliably prepare the computational basis states $\ket{0}$ and $\ket{1}$ as well as the magic state $\ket{A}=\frac{1}{\sqrt{2}}(\ket{0}+e^{i\pi /4}\ket{1})$. We also assume that the classical actions of hardware choice involved in adaptation in response to the classical outcomes of intermediate measurements, can be reliably performed.

It is well known \cite{brkit} that the above resources provide universal quantum computation. Indeed in addition to the Clifford gates it suffices to be able to implement the gate $T= \rm{diag} ( 1\, , \, e^{i\pi /4})$, which can be achieved by an adaptive Clifford process called ``the $T$-gadget'', consuming one copy of $\ket{A}$ as follows: to apply $T$ to line $k$ using $\ket{A}$ on line $a$, first apply Clifford $CX_{ka}$ to those two lines, measure line $a$ to obtain measurement outcome $m$, and then apply the Clifford gate $S^m = \rm{diag} (1\, , \, i)^m$ to line $k$. This results in $T$ being applied on line $k$ (up to an irrelevant overall phase).

In contrast to general adaptive Clifford circuits, in which intermediate measurement outcomes may affect the choice of many subsequent gates, with varying probabilities, giving rise to generally different overall circuit processes, the use of adaptation in the $T$-gadget has the following notable especially simple features:\\
(T1) the adaptation is always only for just a single 1-qubit Clifford gate ($S^m$), done immediately after the measurement;\\
(T2) the measured qubit is immediately discarded upon measurement and its post-measurement state is not further used in the circuit;\\
(T3) the probabilities of measurement outcomes $m=0,1$ are always $1/2$;\\
(T4) the overall computational process is always the same viz. to apply the 1-qubit gate $T$ on a desired qubit line.

For the record, we also recall the following easily proved result (cf  \cite{jvdn}):\\
Let $C$ be any {\em non}-adaptive Clifford circuit i.e. any circuit of unitary Clifford gates also allowing intermediate computational basis measurements, but with choice of Clifford gates not depending on measurement outcomes. Let $C$ be applied to any product state input (not necessarily restricted to computational basis and 1-qubit magic states). Then if any single qubit line is measured at the end of the circuit, the output probabilities can be classically efficiently computed i.e. computed in classical ${\rm poly}(N)$ time, where $N$ is the circuit size. $\Box$

In view of all the above, suppose now that we have solved an instance of a BQP decision task by implementing an adaptive Clifford circuit on our quantum computing device, to obtain output 0 or 1. By further efficient use of the machine and efficient classical computation, we wish to develop confidence in the correctness of the quantum operation of the machine and its classical output for the adaptive choices in the computational run that occurred.

Having run our adaptive Clifford circuit  we record its output as well as the sequence of (adaptively chosen) gates that were actually applied. We will refer to this particular sequence as the `computational run'. To develop confidence in the correctness of its output we next run, polynomially many times, the same sequence of gates non-adaptively (including here also the intermediate measurement operations but ignoring their outcomes). We refer to these runs as `gate test runs'. From the frequencies of their outputs, we obtain an estimate of the output probabilities of this non-adaptive process to within $1/{\rm poly}(n)$ additive error (with probability exponentially close to 1, by the Chernoff bound cf Appendix of \cite{vdn2}). Now this non-adaptive process of Clifford gates and measurements, the same sequence of operations that actually occurred in the computational run, is classically efficiently simulatable. We classically compute its output probabilities and compare them to the experimentally obtained values, verifying that the single actually implemented sequence of operations used in the computational run, provided a true sample of its output distribution to within $1/{\rm poly}(n)$ additive error in the probabilities (assuming that the computing device behaves in the same way for repeated trials of a given process).

The non-adaptive process in the gate test runs differs from the computational run only in that the intermediate measurement results are uniformly randomly varying and unlikely to reproduce those that occurred in the computational run itself. Nevertheless we still develop confidence in the computing device's ability to correctly implement the same sequence of operations that occurred in the computational run, albeit in a slightly different scenario. Indeed since the measured qubit is always disjoint from rest of the computer and never used again, the measurement operation on it can have no effect on the reduced state of the rest of the computer (by the no-signalling principle). As such, the unitary gates of the (adaptive) computational run acted on the same (reduced) input states that occur in the (now classically simulatable,  non-adaptive) gate test runs. Also the computer cannot physically function differently for different intermediate measurement results (assuming suitable non-communication in implementation of local operations, and having the measured qubits reasonably isolated from the other qubits). So although the gate test runs produce generally different intermediate measurement outputs, they may nevertheless still be viewed as providing evidence for validity of the implementation of the quantum operations that actually occurred in the computational run.

In any non-adaptive circuit, if the measured qubit is always discarded upon measurement and never further used, then the final output probabilities will be independent of the intermediate measurement probabilities and in fact, unchanged even if the intermediate measurements are simply omitted i.e. not performed at all. This follows immediately from the fact noted above, that the measurement operation (or even its omission) has no effect on the reduced state of all other qubit lines. Thus the outputs of the gate test runs above provide only limited confirmation of the correctness of the intermediate measurements, in particular of their output probabilities which are required to be (suitably close to) half. 

In the adaptive process, it is important that the different gate sequences (assumed now to be sufficiently faithfully implemented themselves) are chosen with their correct respective probabilities, for a single run of the adaptive process to represent a valid sampling of the desired BQP problem's solution. We can develop further confidence in the device's correctness of its operation in this respect as follows. We consider the initial part of the circuit up to the first $T$-gadget and run it polynomially many times to estimate the measurement probability to within a $1/{\rm poly}(n)$ additive error, and verify that it is within (say) $O(1/t^2)$ of the value half, where $t=O({\rm poly}(n))$  is the number of $T$-gadgets in the circuit. We could further develop confidence in the post-measurement state of the unmeasured qubits by similarly measuring a constant number of them in our polynomially many runs, and again compare frequencies with classically computed ideal values (since output probabilities for a constant number of output lines remains classically efficiently simulatable). Being adequately satisfied with the first $T$-gadget's measurement operation, we apply the same process to the second $T$-gadget, while now treating the first $T$-gadget's measurement output non-adaptively and using the gate sequence from the computational run up to the second $T$-gadget. Similarly we work through all the $T$-gadgets in order. We refer to such test runs as `measurement test runs'.

The theoretical probability of any adaptive gate sequence is $1/2^t$ and we have developed confidence that the computing device has selected the gate sequence used in the computational run with probability

\[ \pi =\left( \frac{1}{2}+O(\frac{1}{t^2})\right)^t = \frac{1}{2^t}\left( 1+O(\frac{1}{t^2})\right)^t. \]
Since $(1+\frac{1}{m^2})^m \rightarrow 1$ as $m\rightarrow \infty$ we see that we can thus (with polynomially bounded computing effort) confirm that $|\pi -\frac{1}{2^t}| <\frac{\epsilon}{2^t}$ for any chosen constant $\epsilon >0$ and all sufficiently large $t$.

Finally let $p_{out}$ be the true theoretical probability of output 0 in the adaptive Clifford process, and let $p^{(j)}_{out}$, for $j=1, \ldots ,2^t$, be the corresponding output probability for the $j^{\rm th}$ adapted gate sequence. (In fact by feature (T4) these are independent of $j$). Also let $\pi^{(j)}$ be the true theoretical probability that the $j^{\rm th}$ gate sequence occurs in an adaptive run. (In fact we have $\pi^{(j)}=1/2^t$ here). Then
\[ p_{out}= \sum_{j=1}^{2^t} p_{out}^{(j)} \pi^{(j)}. \]
Let $j_0$ be the label of the adaptive sequence that was actually used in the computational run.
With polynomially bounded quantum and classical computational resources we have developed confidence that:\begin{itemize}
\item[(i)] (from gate test runs) the output probability $\tilde{p}_{out}^{(j_0)}$ of the implemented computational run is within additive error $\eta = 1/{\rm poly}(n)$ of its theoretical value $p_{out}^{(j_0)}$.

\item[(ii)] (from measurement test runs) the gate sequence labelled $j_0$ has been chosen by the device with probability $\tilde{\pi}^{(j_0)}$ that is within $\epsilon /2^t$ of its theoretical value $\pi^{(j_0)} = 1/2^t$.\end{itemize}
(i) and (ii) then imply that (with suitably chosen $\eta$ and $\epsilon$) our quantum computing device has provided a sample of a probability distribution that is within any desired $\epsilon'>0$ of the theoretical distribution $\{ p_{out}, 1-p_{out}\}$, so its output is then (within the requirements of the bounded error condition) the solution to our BQP decision problem e.g. we could assume without loss of generality that the BQP algorithm used has bounded error margin $|p_{out}-1/2| >0.49$, and choose $\epsilon$ and $\eta$ to   provide $\epsilon'<0.01$, to then establish confidence that the machine's output is the correct answer to the decision problem with probability at least 0.98.

\section{Conclusion}
In common scientific practice, experimental investigations are generally not free of recognised loopholes and are often based on accepted prior assumptions e.g. early Bell inequality tests, despite being open to fundamental loopholes, were nevertheless widely celebrated as highly significant and groundbreaking. Similarly, to accept correctness of functioning of any newly developed computer, the manufacturer generally conducts a series of limited tests that are able to be confirmed within prior trusted technology, and then we are usually content to go on to use the computer in a wide variety of situations, unaccompanied by further targeted testing. 

In contrast to these attitudes, for quantum computing previously proposed verification schemes have generally been required to provide essentially completely unconditional confidence in correctness of the output, but at the same time this supremely strong level of confidence applies only to the single interactive verification run and to no other use of the computer. Our proposed verification scheme on the other hand, is intended to have a different emphasis, more in the spirit of common scientific practice as above. It does not provide unconditional confidence (indeed some pertinent issues have been mentioned in our exposition of the scheme) but instead it is directed towards testing and developing confidence in the validity of operation of the computing device, directly on a desired quantum algorithm, and in preparation for a possible wide variety of further situations. The extent to which it is accepted as adequate will depend on further features of the specific device and application at hand, such as its physical architecture, and the prospective level of malicious or adversarial behaviour that may reasonably be expected to occur.

\end{document}